\documentstyle[twoside,fleqn,espcrc2]{article}

\newcommand{\AmS}{{\protect\the\textfont2
  A\kern-.1667em\lower.5ex\hbox{M}\kern-.125emS}}

\title{Black Hole Entropy from Horizon Conformal Field Theory}

\author{S.~Carlip\address{Department of Physics, University of
        California \\ Davis, CA 95616, USA}%
        \thanks{Supported in part by Department of Energy grant
        DE-FG03-91ER40674}
         }
       
\begin{document}

\begin{abstract}
String theory and ``quantum geometry'' have recently offered 
independent statistical mechanical explanations of black hole
thermodynamics.  But these successes raise a new problem: why 
should models with such different microscopic degrees of freedom 
yield identical results?  I propose that the asymptotic behavior of 
the density of states at a black hole horizon may be determined by
an underlying symmetry inherited from classical general relativity,
independent of the details of quantum gravity.  I offer evidence 
that a two-dimensional conformal symmetry at the horizon, with 
a classical central extension, may provide the needed behavior.
\end{abstract}

\maketitle

\section{SOME QUESTIONS}

More than 25 years have now passed since the discovery by Bekenstein
\cite{Beck} and Hawking \cite{Hawk} that black holes are thermal
objects, with characteristic temperatures and entropies.  But while
black hole thermodynamics is by now well established, the underlying 
statistical mechanical explanation remains profoundly mysterious.
Recent partial successes in string theory \cite{string} and the
``quantum geometry'' program \cite{Ash} have only added to the 
problem: we now have several competing microscopic pictures 
of the same phenomena, with no clear way to understand why they 
give identical results.  

The mystery is deepened when we recall that the original analysis
of Bekenstein and Hawking needed none of the details of quantum gravity, 
relying only on semiclassical results that had no obvious connection 
with microscopic degrees of freedom.  This is the problem of 
``universality'':  why should such profoundly different approaches 
all give the same answers for black hole temperature and entropy?

I will orient this presentation around a few fundamental questions.  At 
first sight, some of these questions---although not their answers---are  
obvious, while others may seem more obscure. I hope to show that these 
are ``right'' questions, in that they lead toward a plausible solution to the 
problem of universality in black hole thermodynamics.  The solution I 
suggest is certainly not proven, however, and perhaps at this stage the 
questions are as important as the answers.

The questions are these:
\newcounter{num}
\begin{list}{$\bullet$}{
   \setlength{\itemsep}{-.25ex} \setlength{\leftmargin}{1.1em}}
\item Why do black holes have entropy?
\item Can black hole horizons be treated as boundaries?
\item Why do different approaches to quantum gravity yield the same
 black hole entropy?
\item Can classical symmetries control the density of quantum states?
\item Can two-dimensional conformal field theory be relevant to realistic
 (3+1)-dimensional gravity?
\end{list}

\subsection{Why do black holes have entropy?}

Our starting point is the Bekenstein-Hawking entropy
\begin{equation}
S = {A\over4\hbar G}
\label{a1}
\end{equation}
for a black hole of horizon area $A$.  It is possible, of course, that black
holes are fundamentally unlike any other thermodynamic systems, and 
that black hole entropy is unrelated to any microscopic degrees of freedom.
But if we reject such a radical proposal, then even knowing nothing 
about quantum gravity, we can make some reasonable guesses.

First, the underlying microscopic degrees of freedom must be quantum 
mechanical, since $S$ depends on Planck's constant.  Second, they must be, 
in some sense, gravitational, since $S$ depends on Newton's constant.  It is 
thus reasonable to suppose that they are quantum gravitational, though this
conclusion is not quite necessary---the relevant degrees of freedom could 
conceivably be those of a quantum field theory in a classical gravitational 
background.  Third, the dependence of $S$ on the horizon area suggests 
(though again does not prove) that the degrees of freedom responsible for 
black hole entropy live on or very near the horizon.

At the same time, we know that the relevant degrees of freedom
cannot be the ordinary ``graviton'' degrees of freedom one expects in 
quantum gravity.  As Ba{\~n}ados, Teitelboim, and Zanelli showed
\cite{BTZ}, black holes exist in (2+1)-dimensional general relativity, 
and exhibit the usual thermodynamic behavior.  But in 2+1
dimensions there are no gravitons, and the ordinary ``bulk'' degrees
of freedom are finite in number \cite{book}.   Indeed, on a spatially
compact (2+1)-dimensional manifold, the bulk gravitational degrees 
of freedom are far too few to account for black hole entropy, and we
can obtain the Bekenstein-Hawking entropy from quantum gravity
only if we admit extra ``boundary'' degrees of freedom \cite{Carlip1}.
Such boundary excitations appear naturally in the Chern-Simons
formulation of (2+1)-dimensional gravity as ``would-be gauge
degrees of freedom,'' that is, excitations that would normally be
unphysical, pure gauge configurations but that become
physical in the presence of a boundary \cite{Carlip2,Carlip3}.

In the (2+1)-dimensional theory, there are two obvious candidates
for a ``boundary,'' spatial infinity and the black hole horizon.  The
degrees of freedom at spatial infinity are naturally described by a
Liouville theory \cite{Henneaux}, and it is not obvious that there
are enough of them to account for black hole entropy.  However,
an elegant conformal field theory argument due to Strominger
\cite{Strom} leads to the correct Bekenstein-Hawking formula.
Unfortunately, though, the theory at infinity cannot distinguish 
between a black hole and, for example, a star of the same mass, 
and cannot easily attribute separate entropies to distinct horizons 
in multi-black hole spacetimes.  For that, we need a ``boundary'' 
associated with each horizon.  This leads to the next question:

\subsection{Can black hole horizons be treated as boundaries?}

A black hole horizon is not, of course, a true physical boundary.  A 
freely falling observer can cross a horizon without seeing anything 
special happen; she certainly doesn't fall off the edge of the universe.
To understand the role of a horizon as a boundary, one must think
more carefully about the meaning of a black hole in quantum
gravity.

Suppose we wish to ask a question about a black hole: for example,
what is the spectrum of Hawking radiation?  In semiclassical
gravity, this is easy---we merely choose a black hole metric as
a background and do quantum field theoretical calculations in that 
fixed curved spacetime.  

In full quantum gravity, though, life is not so simple.  The metric is
now a quantum field, and the uncertainty principle forbids us from
fixing its value.  We can at most fix ``half the degrees of freedom''
of the geometry.  How, then, do we know whether there is a black 
hole?

The answer is that we can restrict the metric to one in which a
horizon is present.  A horizon is a codimension one hypersurface,
and we need only ``half the degrees of freedom'' to fix its properties.
This does not, of course, make the horizon a physical boundary,
but it makes it a place at which ``boundary conditions'' are set.  
In a path integral formulation, for instance, we can impose the
existence of a horizon by splitting the manifold into two pieces
along some hypersurface and performing separate path integrals over 
the fields on each piece, with fields restricted at the ``boundary'' 
by the requirement that it be a horizon.  This kind of split path
integral has been studied in detail in 2+1 dimensions \cite{Witten},
where it can be shown that the usual boundary degrees of freedom
appear.  It seems at least plausible that the same should be true
in higher dimensions.

(I should confess that I have been sweeping a rather difficult
problem under the rug.  It is not clear what kind of horizon
one should choose, or what the appropriate boundary conditions 
should be.  Recent work by Ashtekar and collaborators on 
``isolated horizons'' is probably relevant \cite{Ash2}, but these
results have not yet been applied to this question.)

\subsection{Why do different approaches to quantum gravity yield
  the same entropy?}

We next turn to the problem posed at the beginning of this work,
that of universality.  Ten years ago, if someone had asked for a 
statistical mechanical explanation of black hole entropy, the best 
answer would have been, ``We don't know.''  Today we suffer an 
embarrassment of riches---we have several explanations from 
string and D-brane theory \cite{string}, another from the ``quantum 
geometry'' program \cite{Ash}, yet another from Sakharov-style 
induced gravity \cite{Frolov}.  None of these is yet completely 
satisfactory, but all give the right functional dependence  and the 
right order of magnitude for the entropy.  And all agree with the 
original semiclassical results that were obtained without any 
assumptions about quantum gravitational microstates.

In one sense, this agreement is not surprising: any quantum theory
of gravity had better give back the semiclassical results in an 
appropriate limit.  But the quantity we are investigating, the entropy,
is a measure of the density of states, about as quantum mechanical
a quantity as one could hope to find.  Merely pointing to the 
semiclassical results does not explain {\em why\/} the density of 
states behaves as it does.

This problem has not yet been solved.  But perhaps the most 
plausible direction in which to look for a solution is suggested 
by Strominger's recent work \cite{Strom}.  Regardless of the 
details of the degrees of freedom, any quantum theory of gravity 
will inherit from classical general relativity a symmetry group, the 
group of diffeomorphisms.  While the commutators may receive 
quantum corrections of order $\hbar$, we expect the fundamental 
structure to remain.  So perhaps the classical structure of the group 
of diffeomorphisms is sufficient to govern the gross behavior of 
the density of quantum states.

\subsection{Can classical symmetries control the density of quantum
  states?}

Symmetries determine many properties of a quantum theory, 
but one does not ordinarily think of the density of states as being 
one of these properties.  In one large set of examples, however,
the two-dimensional conformal field theories, the symmetry 
group does precisely that.

Consider a conformal field theory on the complex plane.  The
fundamental symmetries are the holomorphic diffeomorphisms 
$z\rightarrow z+\epsilon f(z)$.  If one takes a basis $f_n(z)=z^n$ 
of holomorphic functions and considers the corresponding algebra 
of generators $L_n$ of diffeomorphisms, it is easy to show that 
\cite{CFT}
\begin{equation}
\left[ L_m, L_n \right] = (m-n)L_{m+n} 
  + {c\over12}m(m^2-1)\delta_{m+n} 
\label{b1}
\end{equation}
where the constant $c$, the central charge, is a measure of the
conformal anomaly.  The algebra (\ref{b1}) is known as the 
Virasoro algebra.

As Cardy first showed \cite{Cardy}, the central charge $c$ is nearly 
enough to determine the asymptotic behavior of the density of 
states.  Let $\Delta_0$ be the lowest eigenvalue of the Virasoro
generator $L_0$, that is, the ``energy'' of the ground state, and
denote by $\rho(\Delta)$ the density of eigenstates of $L_0$
with eigenvalue $\Delta$.  Then for large $\Delta$,
\begin{equation}
\rho(\Delta) \sim \exp
   \left\{2\pi\sqrt{c_{\hbox{\scriptsize eff}}\Delta\over6}\right\}
\label{b2}
\end{equation}
where
\begin{equation}
c_{\hbox{\scriptsize eff}} = c - 24\Delta_0 .
\label{b3}
\end{equation}
A careful derivation of the Cardy formula (\ref{b2})--(\ref{b3}) 
is given in reference \cite{Carlip4}.  The proof involves a simple 
use of duality:

Consider a conformal field theory on a cylinder; then analytically 
continue to imaginary time, and compactify time to form a torus.  
In addition to the familiar ``small'' diffeomorphisms, the torus 
admits ``large'' diffeomorphisms, one of which is an exchange of the 
two circumferences.  Under such an exchange, the angular and time 
coordinates are swapped.  Since the theory is conformal, we can 
always rescale to normalize the angular circumference to have length 
1.  The exchange of circumferences is then a map from $t$ to $1/t$, or 
equivalently from energy $E$ to $1/E$.  The high energy states are thus 
connected by symmetry to low energy states, and one can obtain their 
gross properties---in particular, the behavior of the density of high-$E$
states---from knowledge of low-$E$  states.  The dependence of the 
Cardy formula on the central charge appears because the theory is not 
really quite scale-invariant: $c$ is a conformal anomaly, and the 
rescaling of the angular circumference introduces factors of $c$.

The central charge of a conformal field theory ordinarily
arises from operator ordering in the quantization.  But as
Brown and Henneaux have stressed \cite{Brown}, a central charge
can appear {\em classically\/} for a theory on a manifold with boundary.  
In the presence of a boundary, the generators of diffeomorphisms
typically acquire extra ``surface'' terms, which are required in order
that functional derivatives and Poisson brackets be well-defined 
\cite{Regge}.  These surface terms are determined only up to the
addition of constants, but the constants may not be completely
removable; instead, they can appear as central terms in the Poisson
algebra of the generators.

The canonical example of such a classical central charge is
(2+1)-dimensional gravity with a negative cosmological constant
$\Lambda = -1/\ell^2$ \cite{Brown}.  For configurations that are 
asymptotically anti-de Sitter, the algebra of diffeomorphisms 
acquires a surface term at spatial infinity, and the induced algebra 
at the boundary becomes a pair of Virasoro algebras with central 
charges
\begin{equation}
c = {\bar c} = 3\ell/2G .
\label{b4}
\end{equation}
Strominger's key observation \cite{Strom} was that if one takes 
the eigenvalues of $L_0$ and ${\bar L}_0$ that correspond to a 
black hole,
\begin{equation}
M = (L_0 + {\bar L}_0)/\ell, \quad J = L_0 - {\bar L}_0
\label{b5}
\end{equation}
and assumes $\Delta_0=0$, the Cardy formula (\ref{b2})--(\ref{b3})
gives the standard Bekenstein-Hawking entropy (\ref{a1}) for the
(2+1)-dimensional black hole.

Strominger's derivation is an elegant ``existence proof''
for the idea that black hole entropy can be determined by 
classical symmetries.  As a general argument, though, it has
two obvious limitations.  First, it uses features peculiar to 
2+1 dimensions.  In particular, the relevant Virasoro algebras 
have a natural geometrical meaning: they are the symmetries 
of the two-dimensional boundary of three-dimensional adS 
space.  While many of the black holes in string theory have 
near-horizon geometries that resemble the (2+1)-dimensional 
black hole \cite{Carlip4}, others do not, so this limitation is a 
serious one.

Second, since these Virasoro algebras appear at spatial infinity,
they cannot in themselves detect the details of the interior
geometry.  For example, multi-black hole solutions ought to 
have a distinct entropy attributed to each horizon, but an
asymptotic algebra at infinity can only determine the total
entropy.  Indeed, the classical central charge at infinity cannot 
tell whether the configuration is a black hole or a star.

We are thus led to look for generalizations of Strominger's
approach to higher dimensions and to boundary terms at
individual black hole horizons.  It seems natural to start by
looking for higher-dimensional generalizations of the Cardy
formula.  Unfortunately, no such generalizations are known,
and the derivation of equations (\ref{b2})--(\ref{b3}) does
not naturally extend to more than two dimensions.  We are
thus led to our next question:

\subsection{Can two-dimensional conformal field theory be 
relevant to realistic (3+1)-dimensional gravity?}

We have one example in which symmetries can control the 
density of quantum states: conformal field theory in two dimensions.  
For this fact to be relevant to realistic (3+1)-dimensional 
gravity, we must argue that some two-dimensional submanifold
of spacetime plays a special role in black hole thermodynamics.  
This is in fact the case: in the semiclassical approach to black hole 
thermodynamics in any dimension, all of the interesting physics 
takes place in the ``$r$--$t$ plane.''

Let us consider for simplicity an $n$-dimensional Schwarzschild 
black hole, analytically continued to ``Euclidean'' signature.   (The 
generalization to more complicated black holes is fairly
straightforward.)  Near the horizon, the metric takes the form
\begin{equation}
ds^2 \approx {r_+\over r-r_+}dr^2 + {r-r_+\over r_+}d\tau^2
   + r_+^2d\Omega^2
\label{c1}
\end{equation}
where the horizon is located at $r=r_+$.  It is well known that
the Hawking temperature can be obtained by demanding that
there be no conical singularity in the $r$--$\tau$ plane.
Indeed, if we transform to new coordinates
\begin{equation}
R = 2\sqrt{r_+(r-r_+)} , \quad {\hat\tau} = {\tau\over2r_+}
\label{c2}
\end{equation}
the metric (\ref{c1}) becomes
\begin{equation}
ds^2 \approx dR^2 + R^2d{\hat\tau}^2 + r_+^2d\Omega^2 .
\label{c3}
\end{equation}
Smoothness at $R=0$ then requires that $\hat\tau$ be an ordinary
angular coordinate with period $2\pi$, i.e., that $\tau$ have a period 
$\beta = 4\pi r_+$, the correct inverse Hawking temperature.  Note
that this argument depended only on the geometry of the $r$--$\tau$ 
plane near the horizon, and not on the angular coordinates. 

What is less well known is that the semiclassical computation of
the entropy also depends only on the near-horizon geometry of the 
$r$--$\tau$ plane.  If one chooses boundary conditions appropriate
for the microcanonical ensemble, the classical action of general
relativity reduces to the Einstein-Hilbert action for a ``cylinder''
$\Delta_\epsilon\times S^{n-2}$, where $\Delta_\epsilon$ is a disk 
of radius $\epsilon$ around the point $r=r_+$ in the $r$--$\tau$ 
plane \cite{BTZ2}.  It may be shown that the action factorizes,
becoming
\begin{equation}
\lim_{\epsilon\rightarrow0} I = {1\over 8\pi G}\chi(\Delta_\epsilon)
   \times\hbox{Vol}(S^{n-2}) 
\label{c4}
\end{equation}
where $\chi(\Delta_\epsilon)$ is the Euler characteristic of
$\Delta_\epsilon$.  

In the standard Euclidean path integral approach to black hole 
thermodynamics, the action $I/\hbar$, evaluated with appropriate 
boundary conditions,  gives the leading order contribution to the entropy.  
It is evident from equation (\ref{c4}) that the ``transverse'' factor 
$\hbox{Vol}(S^{n-2})$ is needed to obtain the correct  entropy---it 
gives the factor of area in equation (\ref{a1}).  But this term is also 
nondynamical, merely providing  a fixed multiplicative factor; it is 
the topology of the $r$--$\tau$ plane that distinguishes the black 
hole configuration from any other.

To better understand the relevant conformal field theory, it is useful 
to combine the coordinates $R$ and $\hat\tau$ in equation (\ref{c3})
into a complex coordinate $z=Re^{i\hat\tau}$:
\begin{equation}
z \sim \exp\left\{ {1\over2r_+}
   \left[ i\tau + r_+\ln\left({r\over r_+}-1\right)\right]\right\} .
\label{c5}
\end{equation}
Continuing back to Lorentzian signature, we see, perhaps not
surprisingly, that ``holomorphic'' and ``antiholomorphic'' functions
correspond to functions of $t\pm r_*$, where $r_*$ is the usual
``tortoise coordinate.''

These results suggest two possible strategies for further investigating 
the statistical mechanics of black hole entropy.  We can try to 
dimensionally reduce general relativity to two dimensions in the 
vicinity of a black hole horizon, and see whether we can identify
a conformal field theory and determine the central charge.
Alternatively, we can look at the Poisson algebra of the generators
of diffeomorphism and see whether an appropriate subgroup of
transformations of the $r$--$t$ plane acquires a central charge.

The first of these strategies has been explored by Solodukhin
\cite{Solo}, who has shown that the near-horizon dimensional 
reduction of general relativity leads to a Liouville theory with a
calculable central charge.  While there is still some uncertainty
about the proper choice of the eigenvalue $\Delta$ in the Cardy
formula, it seems likely that the symmetries yield the correct 
Bekenstein-Hawking entropy.  The second strategy is my next 
topic.

\section{ENTROPY IN ANY DIMENSION}

The arguments of the preceding section have led us to a possible strategy
for understanding the universal nature of the Bekenstein-Hawking
entropy.  We begin with classical general relativity on a manifold with 
boundary,  imposing boundary conditions to ensure that the boundary 
is a black hole horizon.  We next investigate the classical Poisson algebra 
of the generators of diffeomorphisms on this manifold, 
concentrating in particular on the subalgebra of diffeomorphisms of 
the $r$--$t$ plane.  We expect this subalgebra to acquire a classical
central extension, with a computable central charge and with some
eigenvalue $\Delta$ of $L_0$ associated with the black hole.  We 
then see whether these values of $c$ and $\Delta$ yield, via the Cardy 
formula, the correct asymptotic behavior of the density of states.  If 
they do, we will have demonstrated that the Bekenstein-Hawking 
entropy is indeed governed by symmetries, independent of the finer 
details of quantum gravity.

The exploration of this strategy was begun in references \cite{Carlip5} 
and \cite{Carlip6}.  While conclusive answers have not yet been obtained, 
the results so far suggest that the program may succeed.  I will now briefly 
summarize the progress.

\subsection{Central terms}

In classical general relativity, the ``gauge transformations''  have 
generators $H[\xi]$ that can be written as integrals of the canonical
constraints over a spacelike hypersurface.  For a spatially closed 
manifold, these generators obey the standard Poisson algebra 
$\{H[\xi_1],H[\xi_2]\} = H[\{\xi_1\xi_2\}]$, where the bracket 
$\{\xi_1\xi_2\}$ is the usual Lie bracket for vector fields, or the 
closely related ``surface deformation'' bracket of the canonical
algebra \cite{Brown}.  As DeWitt \cite{DeWitt} and Regge and 
Teitelboim \cite{Regge} noted, however, the presence of a 
boundary can alter the generators $H[\xi]$: in order for their
functional derivatives and Poisson brackets to be well-defined, 
one must add surface terms $J[\xi]$, whose exact form depends 
on the choice of boundary conditions.
 
The presence of such surface terms can, in turn, affect the
algebra of constraints.  If one writes $H[\xi]+J[\xi]=L[\xi]$,
one generically finds an algebra of the form
\begin{equation}
\left\{ L[\xi_1], L[\xi_2]\right\} = L[\{\xi_1,\xi_2\}] + K[\xi_1,\xi_2]
\label{d1}
\end{equation}
where the central term $K[\xi_1,\xi_2]$ depends on the metric only
through its fixed boundary values.  As Brown and Henneaux pointed out
\cite{Brown}, one can evaluate this central extension by looking at the 
surface terms in the variation $\delta_{\xi_1}H[\xi_2]$, where
$\delta_{\xi_1}$ means the variation corresponding to a diffeomorphism
parametrized by the vector $\xi_1$.  Indeed, since such a variation is
generated by $L[\xi_1]$, it is clearly related to the brackets (\ref{d1}).  
A general expression for these brackets, derived using Wald's covariant 
``Noether charge'' formalism \cite{Wald}, is given in reference 
\cite{Carlip6}.

\subsection{Boundary conditions}

To apply these techniques to a black hole, we must next
decide how to impose boundary conditions that imply the presence of
 a horizon.  This is perhaps the most delicate aspect of the program,
and it is not yet fully understood.  

One possible procedure \cite{Carlip5} is to look at the functional 
form of the metric of a genuine black hole in some chosen coordinate 
system near its horizon, and to restrict oneself to metrics that 
approach this form near the boundary.  This seems fairly straightforward, 
but it may be too coordinate-dependent, and it is not obvious how 
fast one should require metric components to approach their boundary 
values.  A second procedure \cite{Carlip6} is to impose the existence 
of a local Killing horizon in the vicinity of the boundary.  This has the 
advantage of covariance, but it is probably too restrictive for a dynamical 
black hole, and it is again unclear how fast the geometry should approach 
the desired boundary values.   

Each of these choices of boundary conditions leads to the constraint 
algebra (\ref{d1}), with a nonvanishing central term 
$K[\xi_1,\xi_2]$.  Moreover, when restricted to diffeomorphisms 
of the $r$--$t$ plane, this algebra reduces to something that is
almost a Virasoro algebra.  In the covariant phase space
approach of reference \cite{Carlip6}, for example, one finds a
central term of the form
\begin{equation}
K[\xi_1,\xi_2] \sim \int_{\cal H} {\hat\epsilon}
   \left( D\xi_1 D^2\xi_2 - D\xi_2 D^2\xi_1 \right)
\label{d2}
\end{equation}
where $\cal H$ is a constant-time cross section of the horizon,
$\hat\epsilon$ is the induced volume element on $\cal H$, and $D
= \partial/\partial v$ is a ``time'' derivative, that is, a derivative along 
the orbit of the Killing vector at the horizon.

If this expression included an integration over the time parameter  
$v$, equation (\ref{d2}) would be precisely the ordinary central term 
for a Virasoro algebra, and Fourier decomposition would reproduce 
equation (\ref{b1}).  Unfortunately, no such $v$ integration is
present.  This mismatch of integrations was first noted by Cadoni 
and Mignemi \cite{Cadoni} in the study of  symmetry algebras in 
(1+1)-dimensional asymptotically anti-de Sitter spacetimes.  They 
suggested that one could define new ``time-averaged'' generators, 
which would then obey the usual Virasoro algebra.  Alternatively, in 
more than 1+1 dimensions one can add extra ``angular'' dependence 
to the $r$--$t$ diffeomorphisms in such a way that a conventional
Virasoro algebra is recovered.  It was argued in reference \cite{Carlip6}
that such angular dependence may be needed to have a well-defined
Hamiltonian.  However, there is clearly more to be understood.

\subsection{Entropy and the Cardy formula}

Let us set aside this problem for the moment, and suppose that we have 
included a suitable angular dependence or an extra integration over $v$ 
in order to obtain a standard Virasoro algebra.  The resulting central
charge can then be computed as in reference \cite{Brown}, and the 
application of the Cardy formula (\ref{b2})--(\ref{b3}) is straightforward.  
One finds that both the boundary conditions of reference \cite{Carlip5} 
and those of reference \cite{Carlip6} yield the standard 
Bekenstein-Hawking entropy (\ref{a1}).

Although this analysis was carried out for ordinary black holes in general 
relativity, it can be easily extended to a number of other interesting problems.  
The same type of argument yields the correct entropy for cosmological
horizons \cite{Lin}, and probably for ``Misner strings'' in Taub-NUT
and Taub-bolt spaces \cite{Carlip6}.  The analysis can be easily
applied to dilaton gravity as well, where it again gives the correct entropy.

Several problems remain.  The first two, which are probably related, are 
to find the right general boundary conditions and to understand the extra
integration required in equation (\ref{d2}).  

The third is conceptually more difficult.  The approach to black hole
entropy described here is nondynamical: by imposing boundary conditions
at a horizon, we are fixing the characteristics of that horizon, 
effectively forbidding such processes as Hawking radiation.  This is not
a terrible thing if we are only interested in equilibrium thermodynamics;
the Bekenstein-Hawking entropy, for example, is presumably ``really''
the entropy of a black hole in equilibrium with its Hawking radiation.
To fully understand the quantum dynamics of a black hole, however,
one needs a more general approach, in which boundary conditions are
strong enough to determine that a black hole is present, but flexible enough
to allow that black hole to evolve with time.  I leave this as a challenge 
for the future.

\end{document}